# Workload Classification & Software Energy Measurement for Efficient Scheduling on Private Cloud Platforms

James W. Smith
School of Computer Science, University of St Andrews
St Andrews, Fife, KY16 9SX
+44 (0) 1334 463253
james.w.smith@st-andrews.ac.uk

Ian Sommerville
School of Computer Science, University of St Andrews
St Andrews, Fife, KY16 9SX
+44 (0) 1334 463253
ian.sommerville@st-andrews.ac.uk

## ABSTRACT
At present there are a number of barriers to creating an energy efficient workload scheduler for a Private Cloud based data center. Firstly, the relationship between different workloads and power consumption must be investigated. Secondly, current hardware-based solutions to providing energy usage statistics are unsuitable in warehouse scale data centers where low cost and scalability are desirable properties. In this paper we discuss the effect of different workloads on server power consumption in a Private Cloud platform. We display a noticeable difference in energy consumption when servers are given tasks that dominate various resources (CPU, Memory, Hard Disk and Network). We then use this insight to develop *CloudMonitor*, a software utility that is capable of >95% accurate power predictions from monitoring resource consumption of workloads, after a "training phase" in which a dynamic power model is developed.

## Categories and Subject Descriptors
C.4 [**Performance of Systems**]: Measurement Techniques; K.6.2 [**Computer Milieux**]: Management of Computing and Information Systems – *Installation Management*

## 1. INTRODUCTION
Data centres can have a massive impact on the IT sector's energy consumption and carbon emissions. The United States Environmental Protection Agency (EPA) highlighted some key issues with regards to computing data centres in their 2007 report:

[1]

- U.S. data centers consumed 1.5% of the total electricity generated in the whole country that year, the equivalent of the combined consumption of 5.8 million average U.S. households or a cost of roughly $4.5 billion dollars.

- The amount of total energy consumption in this area has doubled in the period 2000-2006 and similar growth is expected in the coming years.

- Only 50% of the energy consumed by data centers can be attributed to the useful work done by the computing servers. The other 50% is expended on infrastructure needs such as climate control, lighting, security, etc.

Corporate and government data centres are being transformed into so-called "Private Cloud" systems. Early research into enterprise adoption of Cloud Computing has shown a reluctance to embrace public cloud services despite their potential economic benefits [2]. Issues such as security, privacy, and even socio-technical factors like employee concerns have so far restricted the mass adoption of public Cloud services amongst enterprises. There is lack of support for decisions makers and even the economic benefits are complex, leading to difficult decisions and hesitation. Businesses would rather employ their own people to administer hardware they own, on premises, with controlled access and established security procedures than risk their future using an unfamiliar and evolving technology

Private clouds comprise a shared pool of configurable computing resources that may be rapidly provisioned, released, scaled up or down in proportion to workload. Examples of these types of system include VMware's vSphere and the open source project Eucalyptus that attempts to mimic Amazon Web Services EC2 functionality and provides the same interface. These systems provide infrastructure as a service (*IaaS*), or virtual machines on-demand.

By introducing a virtualised environment to a local data center and accepting the inevitable performance degradation it is possible to introduce a new level of flexibility especially with regards to efficient scheduling of different workloads.

Facebook, Inc. announced in April 2011 that their new data center in Oregon, USA has a power Usage Effectiveness (PUE) rating of 1.07 [3], meaning that 1 in every 1.07 Watts of electrical power entering their plant goes directly to powering servers doing useful work. The margins for improving power consumption by decreasing the amount spent on auxiliary work and other mechanical areas (such as cooling) are increasingly slim. The largest remaining potential for reducing power costs is increasing the efficiency of the IT systems themselves.

Before we can begin to improve the efficiency of a system, we first have to understand it. One of the main issues relating to Energy Efficiency of any system, not only IT, is a lack of monitoring. An NCC UK report estimated that only 13.4% of organisations monitor power consumption [4]. Clearly, introducing monitoring will not only informing users and system administrators about their consumption levels, but will also have the *potential* to impact how these systems are operated by allowing management polices that adapt to energy usage. Energy monitoring is also expensive; requiring additional hardware and new monitoring frameworks such as those proposed by [5].

However, energy usage is difficult to monitor. Traditionally it requires the use of dedicated hardware that will monitor the electric cables supplying the machine.

Investigations are underway at the Cloud Computing Co-Laboratory in St Andrews (StACC) to investigate the connection between workload and power consumed by computing devices. We approach this challenge by offering a Eucalyptus Private Cloud service to users and monitor the subsequent power and resource usage of the provisioned virtual machines and physical servers.

In the case of the StACC Private Cloud, hardware power metering is done through a power Distribution Unit (PDU) which provides per socket (and therefore per machine) measurement.

In a cloud system, one of the main desirable properties is scalability something that is not possible through the use of dedicated energy monitoring devices. Therefore, we seek an alternative means of calculating the energy consumption through software. Initially, such a tool would be initially trained against a PDU, to allow it to ascertain the connections between particular resource consumption and total energy used. This training phase would allow automated generation of a suitable power model to express how the different resource consumptions affect the total power used. This paper presents the initial development of such a tool; *CloudMonitor*, an automated, scalable Energy and resource reporting utility.

This tool is the first step in a process to provide a modified Eucalyptus Private Cloud system that balances resource requirements to provide optimal energy consumption across the entire system. Virtual Machines in the Private Cloud will be assigned a profile that identifies which particular resource dominates its needs (from CPU, Memory, Disk, Network). Those VMs will then be scheduled across the physical machines in a manner that aims to improve the overall efficiency of the system.

In this paper we first present a brief review of relevant current work in the areas of Energy Efficient and Cloud Computing, we will then progress to present an experiment showing the different affects of resource Utilisation on power Consumption, before finally presenting and evaluating a software solution for predicting power usage.

## 2. RELATED WORK

The main difficulty in successfully monitoring a Cloud System in order to improve its energy efficiency lies in the use of hardware PDU to give accurate figures on power usage.

An alternative solution is required, as noted by Stoess et al [6] who discussed the possibility of energy measurement for Virtual Machines, which clearly cannot be connected to a hardware measurement device.

Stroess et al proposed that power usage could be accurately calculated so long as each hardware device reported its power usage in each power mode to its device driver, which in turn would expose that information to the operating system. However, this would require vendor support and is not applicable to systems where their hardware does not support this kind of fine-grain monitoring of individual components.

Snowdon et al [7], Zeng et al [8] discuss approaches to monitoring power for applications, a technique that may be applicable to our Cloud Platform. This involves collecting information in real time about the resource consumed by each application. Snowdon's work assumes that the energy usage of the application is directly related to the amount of time spent as the active thread on the CPU. It does not take into account work done elsewhere such as Buffered IO and multi-core processing. Othewise, Zeng's work relates to Energy budgeting and took data-sheet numbers for several key hardware components.

The most complete work in the area is discussed in the paper "*Virtual Machine power Metering and Provisioning*" by Kansal, et al [9]. The software they present, *Joulemeter*, uses only power models to accurately infer the power consumption of a Virtual Machine. The models are built, or trained, on servers with existing hardware power measurement capability and then are deployed across many of the same type of server without the need for additional hardware measurement.

This ability is obtained through resource monitoring, as the authors explain, "…in principle VM power can be metered by tracking each hardware resource used by a VM and converting the resource usage to power usage based on a power model for the resource." This is a good indicator of where the StACC Private Cloud would need to go to create the same type of measurement for use in effectively scheduling work on the Eucalyptus Private Cloud system.

Bohra & Chaundary in their paper *VMeter* [10], describe a method for predicting the power usage of Virtual Machines by monitoring the consumption of hardware resources on the host server, in particular CPU, cache, DRAM and Hard Drive. Using this method they claim 94% accuracy when compared with against the actual measured power using an externally attached power meter. The accuracy of the power meter they used might be questioned, as it a consumer off-the-shelf device and does not guarantee billing-level accuracy for server environments.

This is an improvement over other approaches to Energy measurement in Cloud systems, reducing cost, increasing reliability and improving scalability, as they are not installing dedicated energy monitoring hardware for all computers in the data center.

Their proposed power model is entirely sensible; it is based upon the linear relationship between the sub-components of the system and groups together hardware resources into CPU-bound (CPU, cache) and IO-Bound (DRAM, HDD).

They decompose the total system power consumption into two major categories: the baseline power consumption (power consumed when the system is on but doing no useful work) and the dynamic power consumption.

Baseline power is given by (notation is preserved from Bohra & Chaundary's [10] work) :

$$P_{\{baseline\}} = \alpha * a_1 + \beta * a_4$$

Total power consumption is given by:

$$P_{\{total\}} = P_{\{baseline\}} + \sum_{1}^{k} P_{\{domain(k)\}}$$

Domain power:

$$P_{\{domain(i)\}} = \alpha \left(a_2 * P_{\{CPU(i)\}} + a_3 * P_{\{cache(i)\}}\right) + \beta \left(a_5 * P_{\{DRAM(i)\}} + a_6 * P_{\{HDD(i)\}}\right)$$

The model sensibly breaks down power consumption per domain to the power calculated for each resource multiplied by some weight to achieve the total power for the system. However, weights $\alpha$, $\beta$, $a_1$, $a_2$, $a_3$ $a_4$, $a_5$ $a_6$ are workload specific and are calculated based on each benchmark programs for all our power model evaluation figures. This appears to be done manually.

Our work has improved this area by automatically analysing resource consumption to predict power usage. This low-cost, scalable and cross platform solution can be rolled out across an entire Cloud Platform without introducing any additional cost.

## 3. WORKLOAD EXAMINATION

### 3.1 Hardware Setup

The StACC Private cloud has the ability to provide per socket (effectively per server) energy consumption information including wattage, current and voltage drawn, using a power Distribution Unit (PDU) to give an accurate picture of power consumed by each physical server.

We use a Rairitain PX-5367 PDU to power the servers in our experiments. It is possible to request power usage information for those servers that are connected to the intelligent PDU by directly querying the factory provided SNMP API.

Simple Network Management Protocol (SNMP) defines a means of exchanging information between one or more network management systems and a number of agents and provides a means for formatting and storing management information [11]. In this case the PDU acts as another agent, providing real time data in response to queries.

Voltage applied, current drawn, active power being used and total watt-hours are among the statistics provided. We use Active power as a snapshot measurement of how much electricity the server is requiring at a given moment. The Rairitain PDU provides refreshed data at 3-second intervals, with those values being the calculated average of fluctuations during the 3-second period.

Our experimental setup used some of the StACC Cloud compute nodes; 2010 Dell PowerEdge R610 servers each with 2x 4 core Intel Xeon (E5620) Processors clocked at 2.40GHz, 16GB of DDR RAM and 146GB SAS 10,000RPM Hard disks. Each was connected to the network via a single gigabit RJ45 link.

Each server was connected to a socket on the Rairitain PX-5367 PDU allowing real time billing level monitoring of power usage.

### 3.2 CloudMonitor System Design

There has been discussion upon the possible types of architecture suitable for Cloud Computing measurement [5] and we broadly follow Yu et al's recommendations. Our system is structured around a centralised data-store that maintains all monitoring information from various agents. The main benefit of this model that agent numbers are unlimited and report their recorded information back to the centralised store that can then compute any required analysis without additional load on the compute servers.

While a single point of failure is normally avoided in Distributed System design, most Cloud platforms do include a single controller that decides where new workloads should be placed. As it is our long-term goal to use this monitoring information to inform these scheduling decisions, it makes sense to put our monitor controller on the same physical machine as the Cloud controller.

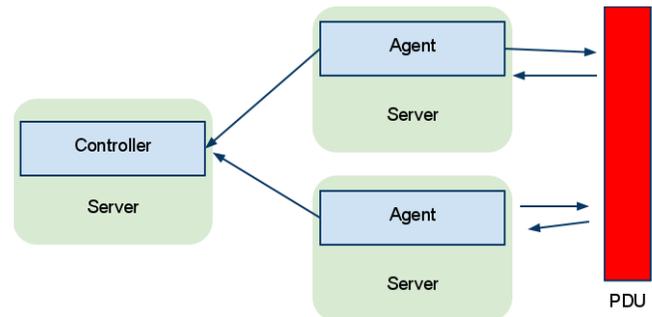

**Figure 1. The *CloudMonitor* Architecture**

Each portable monitor agent sends data back to the control for analysis once every 3 seconds. This data includes Hardware resource usage and potentially corresponding PDU data for the same system. PDU information requests are made from the system being monitored, as this is an option task. The applications are configured on deployment and if the relevant option is selected then the system will request the relevant data from the PDU. There are two main benefits to this approach: firstly that the data being sent to the centralised data store is already packaged with all of the relevant data, and secondly that this approach reduces complexity by allowing the centralised data-store (and eventually, scheduler) to run independent of the current monitoring deployment.

### 3.3 Implementation

The agent programs are written in the Java programming language, to provide portability, allowing deployment on any platform. While Java may not be the language with the lightest footprint, its ease of deployment far out-weighs any performance penalty.

We selected the Hyperic's "System Information Gatherer" (SIGAR) Library to monitor the system sub-components (CPU, RAM, HDD, Network). SIGAR provides a cross-platform API for collecting data on resources and processes. This information is also available directly from most operating systems by recording system events [12], or by accessing hardware performance counters [13]. However, we utilise the SIGAR API as it allows us to develop our system abstracting over the details of the underlying platform. This, coupled with using the Java programming language for the rest of the *CloudMonitor*, provides portability, allowing us to deploy the monitor on Linux, FreeBSD, Windows, Solaris, AIX, HP-UX and Mac OSX across a variety of versions and architectures without changes to the program code.

## 3.4 Experiment

With our test server connected to the power Distribution Unit for Energy monitoring and our *CloudMonitor* utility in place to record resource usage data, we now began generating synthetic computationally diverse workloads using the Phoronix Test Suite[1].

Phoronix is a comprehensive test suite for stressing a system's different sub-components. It is multi-platform, open source and provides a vast library of available tests, broken down by their resource testing objective; for example Processor, Hard Disk, etc. From the available libraries the following tests were chosen:

| resource sub-component | Selected Benchmark |
|---|---|
| CPU | *7zip* File Compression |
| Memory | *gzip* File Compression |
| Hard Disk | AIO-Stress I/O |
| Network | Loopback TCP Net Performance |

**Table 1. Sub-component benchmark selections**

*7zip* is a lightweight standard compression tool, with quite an observed overhead on CPU and required memory. During the normal compression of a large file, CPU usage remains constant at the maximum speed possible, whereas RAM usage fluctuates.

Similarly, *gzip* is also a compression tool, but without the large observed memory and CPU overhead. The performance of *gzip* depends upon primarily the speed at which RAM and cache can be accessed. In the test, *gzip* generates workload by compressing a large 2GB file.

*AIO-Stress* is an asynchronous I/O benchmark created by SuSE. Using a single thread and consistently writing and reading a 1024MB test file to and from the hard disk.

*Loopback TCP Net Performance* uses the system's loopback network adaptor to simulate network traffic in order to benchmark the subsequent machine performance of preparing to transmit and receiving a large amount of data.

Each test was performed in sequence and results are presented in the following section. Every test was performed multiple times and checked for consistency, but for clarity the results presented below represent a selection of tests performed sequentially.

## 3.5 Results

### 3.5.1 7zip – CPU Benchmark

The first graph, Figure 2, shows power consumed by the test machine over the first test; the computationally demanding benchmark *7zip*. In this benchmark, CPU is almost consistent at 100%, but the memory usage varies. Figure 2 shows the power consumption of this test. Notice that the shape of the graph shows that the power fluctuates quite dramatically, the range (or average) power is quite high, regularly over 200W.

The range of the power graph is 100W to 250W, with those being the extremes of observed power states for all experiments. All graphs of the same type are given the same scale, to allow differences to be clearly shown and understood.

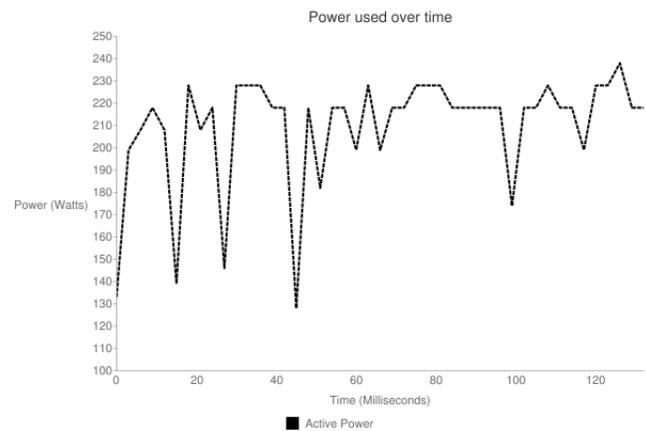

**Figure 2. power over Time for 7zip benchmark**

As we know that CPU is consistent, the question arises as to what causes the power fluctuations to occur. Figure 3 holds the answer. Notice that large amount of change in memory usage over the course of the benchmark. These changes have a significant affect on power usage. There are no or little changes in the use of other resources. This shows that power consumption is not entirely reliant upon CPU alone.

---

[1] http://www.phoronix-test-suite.com/

[2] http://www.cs.st-andrews.ac.uk/stacc

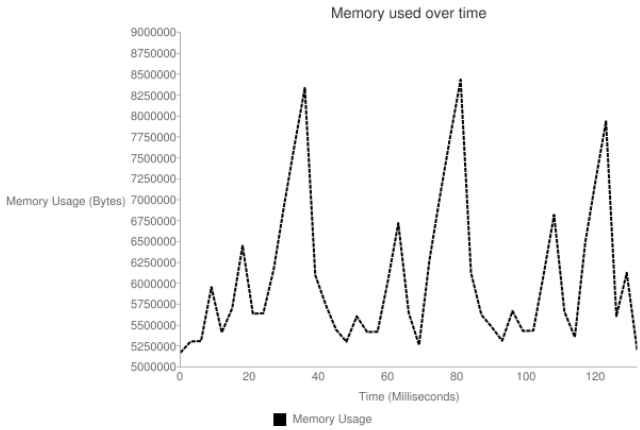

**Figure 3. Memory Usage over Time for 7zip Benchmark**

Figure 4 shows that the CPU usage over the course of this benchmark is mostly constant, confirming that the majority of change can be directly attributed to Memory usage. However the high degree of average power (>200W) can be partly attributed to high CPU utilization, which as Figure 4 shows us is mostly 100% (1.0 of CPU on the graph equates to 100% CPU utilization, 0.5 = 50%, etc).

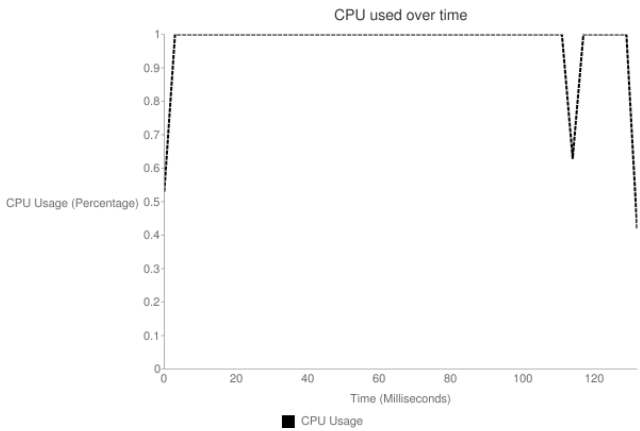

**Figure 4. CPU usage over Time for 7zip Benchmark**

### 3.5.2  gzip – Memory Benchmark

*gzip* is an application that uses a large, consistent amount of memory, but a variable amount of CPU over time (the opposite of 7zip). Figure 5, below, shows that an almost constant amount of energy is used with an average of around 135W – much less than the previous *7zip* benchmark. This shows that two applications, doing much the same task, can have vastly different power consumption.

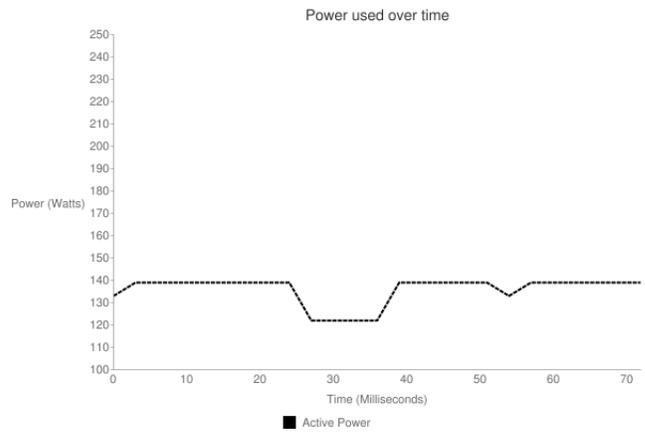

**Figure 5. power over Time for gzip Benchmark**

However, there are fluctuations in the shape of the graph leading us to investigate the cause as the previous culprit; memory is shows to be almost consistent in this test in Figure 6.

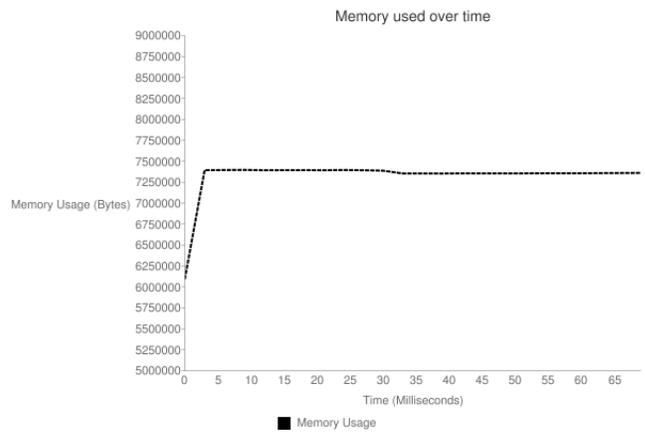

**Figure 6. Memory Usage over time for gzip Benchmark**

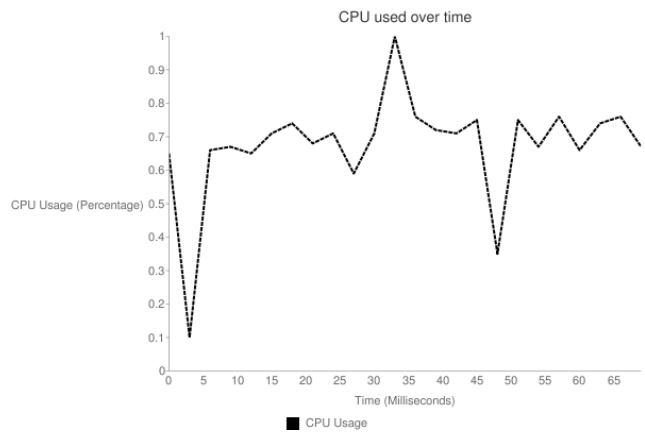

**Figure 7. CPU usage over Time for gzip Benchmark**

And Figure 7 shows CPU performance over the same timeframe. There appears to be some correlation between CPU usage percentage and the amount of power consumed, as no other resources fluctuate their utilization levels over the course of this test. The power graph is a little delayed compared to the CPU, due to the PDU taking an average reading over the last sample interval.

From these results we are beginning to see a relationship develop between different resource utilization levels and the total amount of power consumed.

### 3.5.3 AIO-Stress – Hard Disk Benchmark

In this test, Hard disk usage is almost constant over time, with Memory and CPU fluctuating. Figure 8 shows the power Usage for the duration of the test.

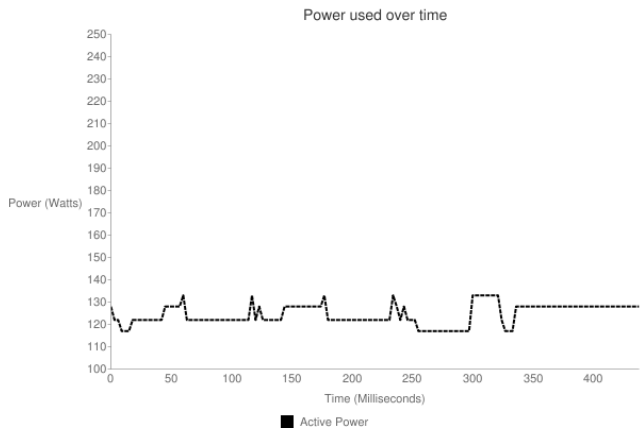

**Figure 8. power Usage over Time for AIO-Stress Benchmark**

power averages a little below level as the previous test, *gzip*, but as Figure 9 shows, CPU usage is much less on average, abit with more surges in use. Further investigation is required to ascertain the cause of these power values.

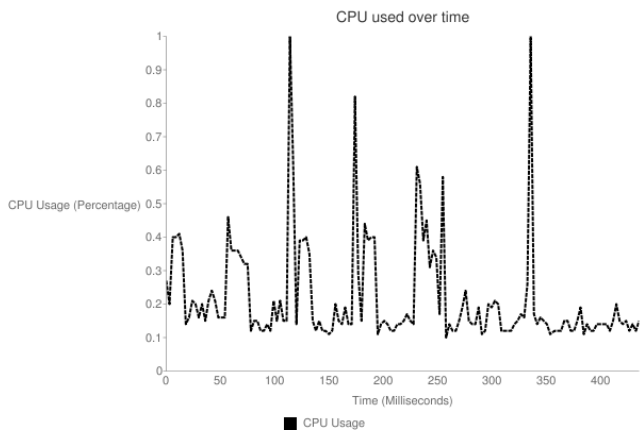

**Figure 9. CPU usage over time for AIO-Stress Benchmark**

Figure 10 shows the Memory usage for this test, but we see that it is a quite low, relatively consistent amount. Further investigation is required.

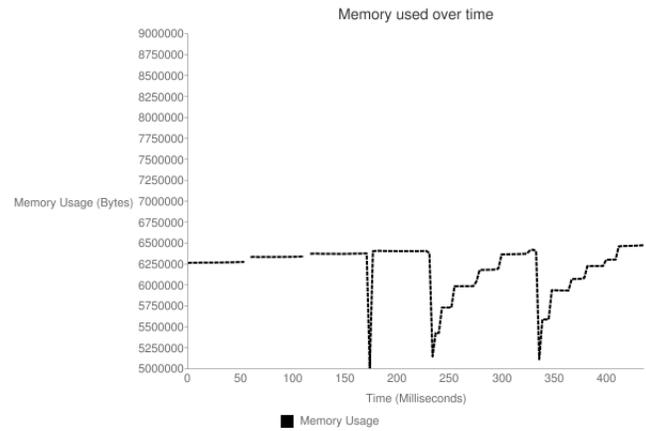

**Figure 10. Memory usage over time for AIO-Stress Benchmark**

Now the Hard Disk performance for that same test is given is Figure 11.

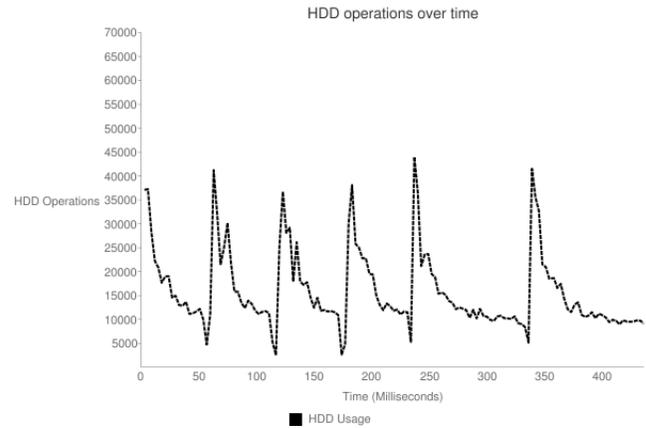

**Figure 11. Hard Disk Operations over Time - AIO Stress**

Figure 11 shows a large number of constant Hard Disk operations for the duration of this benchmark and could provide an answer for the spikes of power usage over the test. This show Hard Disk usage could also be an important property for determining overall power consumption.

### 3.5.4 LoopbackTCP – Network Benchmark
The final test looks at network performance. The power usage for this test is shown in Figure 12.

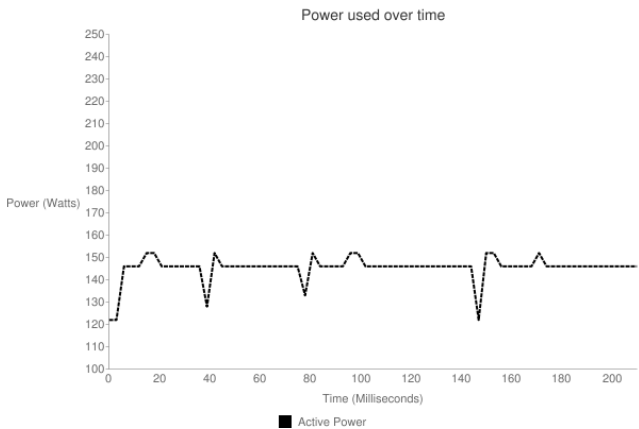

**Figure 12. power over time – Network Benchmark**

Notice here that there is a relatively high level of power consumed. However, the values match none of the earlier tests, leading to question as to how this power consumption graph is achieved.

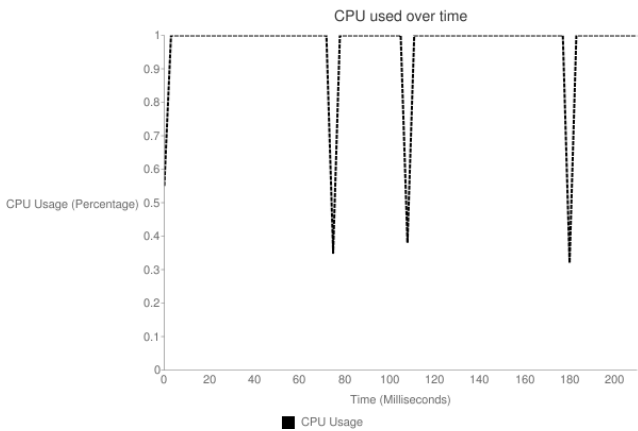

**Figure 13. CPU over Time - Network Benchmark**

Figure 13 Shows that CPU usage is high over the time of this benchmark, a few dips not withstanding. The only previous test to have this level of CPU utilization was *7zip*, but it had a much higher level of Memory usage contributing to a higher overall power consumption. Figure 14 shows that this test did not have the same level of memory usage.

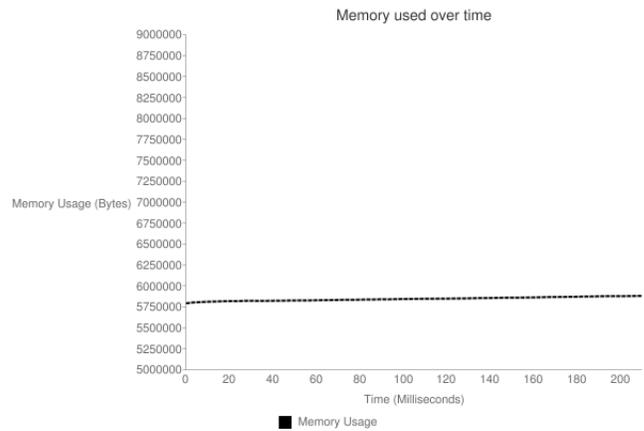

**Figure 14. Memory over Time - Network Benchmark**

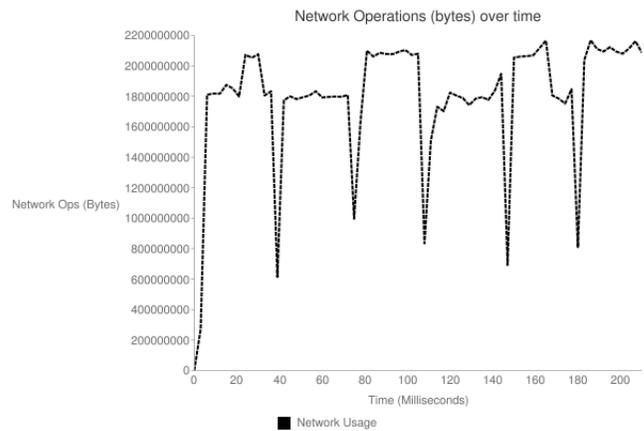

**Figure 15. Net Operations over Time - Network Benchmark**

How then, was the Particular power level shown in Figure 12 reached? Figure 15 shows a vast number of Network operations per sample interval over the course of the benchmark. This explains the particular power model for this test.

### 3.5.5 Summary

Over the course of these tests we can see that there is a significant correlation between workload types and power usage fluctuations. The *7zip* test showed that power Fluctuations are not reliant upon CPU alone, but that paging data in and out of memory at various rates can add a large penalty to the system power consumption.

The next test, *gzip*, performed a similar task of file compression, but by using a constant amount of memory usage and less CPU on average, it was able to require a far less amount of power.

*AIO-Stress* benchmark used less CPU than all over the other tests, and as shown it had the lowest amount of power Consumption. However that level of consumption was not as significantly lower as the reduced amount of CPU and Memory would suggest, leading to the conclusion that the increase in Hard Disk operations offset the savings in the other sub-components.

Finally we saw that under the Network Simulation test *LoopbackTCP* that despite a relatively high level of CPU being used, the lack of Memory usage fluctuations lead to a lower overall power consumption compared to the similarly CPU-demanding *7zip* test. This suggests that some workloads may not have the same affect as others (compare Network to Memory), but each has a significant affect on the total amount of power consumed and this work opens the door to the possibility of energy efficient scheduling of workloads by examining the resource utilization requirements.

## 4. SOFTWARE ENERGY MEASUREMENT

### 4.1 Extending CloudMonitor

Now that workload investigations have opened the possibility of efficient scheduling on Cloud Platforms, we must next satisfy the requirements of low-cost and scalable power metering.

We have extended our tool, *CloudMonitor*, to follow the suggestion of Kansal, et al [9] and infer the power consumption from software alone using computationally generated power models. Our system has been implemented and deployed on a Linux platform using the Java programming language and the previously detailed SIGAR library. This results in a portable utility that is deployable in most IT environments.

The section of the code that directly communicates with the PDU, which we call *EnergyMonitor*, is a separated component from the rest of the program and can easily be replaced with an alternative specifically tailored to any PDU system. Guidelines on how to implement a Monitor for another PDU will be available on the StACC Project website[2].

As in the initial workload investigations, our test server was connected to a PDU to allow power consumption data to be recorded. Our software utility will record the hardware utilization. We are then able to compare the resource utilization information with the PDU data. The data was analysed and multiple linear regression techniques applied to calculate appropriate weights for each hardware sub-component. This "training phase" allowed a suitable power model to be automatically generated.

### 4.2 power Model

We use the Multiple Linear Regression class of *Michael Thomas Flanagan's Java Scientific Library*[3] to accurately assess the coefficients of each set of power and resource measurements. Our power model, based on the work by Bohra & Chaundary [10], is given as follows:

$$P_{\{total\}} = \alpha + (\beta_1 * CPU) + (\beta_2 * RAM) + (\beta_3 * HDD + \beta_4 * Network)$$

This model takes into account each hardware resource that we measure and generates the weights $\alpha$ and $\beta_1, \beta_2, \beta_3, \beta_4$ automatically during the training phase by providing a real world measurement of $P_{\{total\}}$

---
[2] http://www.cs.st-andrews.ac.uk/stacc

[3] http://www.ee.ucl.ac.uk/~mflanaga/java/

As machines in a data center are normally procured in batches, this training phase is only required on one machine per batch. The resulting power model, incorporated into the resource utilization monitor, is able to predict power usage across the remaining servers without the need for dedicated energy monitoring hardware. This not only saves complexity but also operational costs.

### 4.3 Experiments

The same set of experiments as in the initial workload examination were carried out on the same hardware, to allow comparison of actual recorded power usages and those figures generated from the power model. Each of the selected tests were run sequentially with the power Model only evaluated at the end of these computationally diverse workloads.

Once a power model was developed *CloudMonitor* would use the power model to calculate the required Energy usage for the resource statistics of each test and the accuracy of the power model was determined.

### 4.4 Results

For the *7zip* benchmark power measurements are as shown in Figure 16:

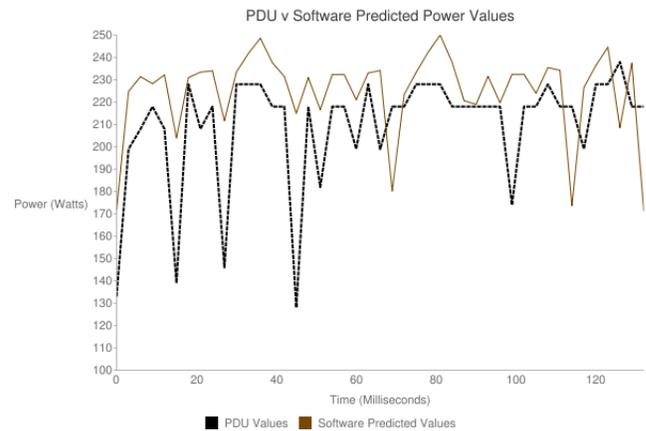

**Figure 16. PDU and Predicted power for 7zip**

The Darker line shows power usage values as recorded by the PDU over the lifetime of the test, the lighter line shows the predicted power usage as calculated by the power Model.

The graph shows that there is strong correlation between the predicted power value and the actual as given by the PDU. The *CloudMonitor* power model even appears to react quicker to fluctuations in the hardware resource usage, where as the PDU takes an average over the 3-second sample interview to produce a smoothing effect.

Tests to calculate the average error affect of the Predicted power Values against the PDU values were run for all of the computationally diverse benchmarks, and over a large sample size (over 291 measurements) the Monitor records is only 4.92% standard error deviation rate, or is >95% accurate, well within suitability requirements.

The following graph is an example of a consistent error rate, in the *gzip* benchmark.

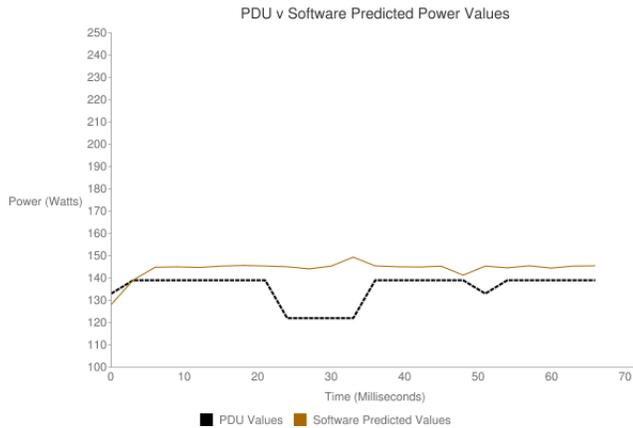

**Figure 17. PDU and Predicted power for gzip**

In Figure 17, the average power from the PDU is set at around 139W, where as the CloudMonitor predicts a value of 145W, an average error over the course of the benchmark of only 4.31%.

This level of error could be considered acceptable for scheduling, particularly given the low costs and scalability of such a software based solution. Local tests have shown that there is potential for dynamic power model generation on only one machine in a batch followed by rollout on all others, but these experiments are still in the initial stages and will be the subject of future work.

# 5. CONCLUSIONS & FUTURE WORK
## 5.1 Conclusions
These results demonstrate that our hypothesis, that different workloads have a noticeable affect on energy consumption and that a low-cost, scalable, cross-platform software solution for energy measurement is possible.

In Section 3.5 we showed that different workloads can have a significant, yet predictable, affect on the total power consumption of a system. This opens the door to the possibility of scheduling workload mixes for optimal performance and energy efficiency. Additional research will now be required to investigate optimal mixes of workload types, with the view to scheduling in such a manner.

In Section 4 we examined the possibility of predicting power usage using only software-based resource Monitoring and multiple linear regression data analysis tools. The result proved that our utility was able to predict power levels to within a 5% standard error margin, an acceptable level for scheduling, given the benefits of a software-only solution.

Currently resource information is gathered as 3 second "snapshots", with no regard to utilization changes within that sampling rate. Further work on the *CloudMonitor* could extend the software to take into account changes over this period and would likely prove to introduce a "smoothed" affect for predicted power, much like the current sampling model of the PDU.

The next step is to apply our techniques to other models of server, rather than the single brand of machines configured in exactly the same way. We have such machines in the StACC Private Cloud and will be rolling out experiments shortly.

## 5.2 Applications
On-going work within StACC is focused on developing a system that will take into consideration two factors, the total energy consumed by the system and the performance experienced by each Virtual Machine. Taking cue from the insights gained by monitoring the StACC Private Cloud, work has begun to create profiles representing workloads in Virtual Machines. At the moment, these workloads are specified by their intensity in certain resources (such as CPU, Disk, Memory and Network).

It is intuitive that pairing two CPU intensive Virtual Machines on the same physical server will have a detrimental effect on their respective performance. Instead, it seems wiser to separate Virtual Machines that are intensive in the same resource, and attempt to pair them instead with VMs that do not consume a large amount of that resource, i.e. pairing a CPU intensive VM with a Disk intensive VM on the same physical server.

Observations from the StACC have shown that VMs are commonly doing only a single task during their lifetimes; predominant use of a particular compute resource will emerge. Our monitoring has shown that few applications are equally hungry in multiple resources. The profile will identify the type of resource need for a each VM in the form of meta-data attached to the instance creation request.

Users will be able to specify how they intend their VM to be used by attaching an easily editable XML file to the instantiation request. If the user is unsure how their application will perform, then they will able to run it alongside a resource-monitoring framework by specifying a flag on run time. The framework will display the current appropriate profile for that VM in the instance description available on the StACC Private Cloud.

Using these profiles, the scheduler on the Eucalyptus open source cloud platform (the software that powers the StACC Public Cloud) will be re-written to allocate Virtual Machines according to the current state of the system and the VM profile's indicated needs.

It is entirely possible with this model for users to lie when specifying their Virtual Machine profile, by labelling a CPU intensive VM as Network intensive for example. This will have a detrimental effect on the system performance, but we believe that the users of Private Cloud systems are concerned with increasing performance and lowering energy consumption figures so will likely be truthful in using the system correctly if it provides some benefit to their goals.

This system is designed as an off-line algorithm where a decision as to where the Virtual Machine should be placed is made once and not changed throughout the lifetime of the Virtual Machine. This obviously does not take action when changes in the VM behaviour make the initially assigned profile irrelevant. In this case, which from our observations we believe is rare due to the single tasks of VMs and the predictable nature of those tasks, the system will continue to operate as normal.

The benefit of this approach is that the penalty of VM teleportation is not incurred; meaning little resources are wasted doing useless, administrative work. Teleportation of executing virtual machines is also a non-trivial task, requiring temporarily quiescence VMs in order to achieve state consistency throughout the movement process.

Work in this area will now continue, with the aim to producing an optimal mix of Workloads to reduce system energy consumption while preserving performance.

## 6. ACKNOWLEDGMENTS

The authors would like to thank their colleagues at the University of St Andrews and LSCITS (www.lscits.org), especially Angus Macdonald and David Greenwood, for their invaluable advice and guidance during this work. We thank the Scottish Informatics and Computer Science Alliance (SICSA) and the EPSRC (grant numbers EP/H042644/1 and EP/F001096/1) for funding this work.

## 7. REFERENCES

[1] EPA "*Report to Congress on Server and Data Center Energy Efficiency*", 2007, Public Law, Vol 109, p. 109-431

[2] Khajeh-Hosseini, A., Greenwood, D., Smith, J. W. and Sommerville, I. (2011), The Cloud Adoption Toolkit: supporting cloud adoption decisions in the enterprise. Software: Practice and Experience, 41: n/a. doi: 10.1002/spe.1072

[3] Facebook Launches Open Compute Project; Press Release: https://www.facebook.com/press/releases.php?p=214173

[4] EDS and NCC, The Green IT paradox: Results of the NCC Rapid Survey, EDS; NCC, 2009

[5] Yu, Yi, and Saleem Bhatti. "Energy Measurement for the Cloud." In *International Symposium on Parallel and Distributed Processing with Applications*, 619–624. IEEE, 2010. http://www.computer.org/portal/web/csdl/doi/10.1109/ISPA.2010.29.

[6] J. Stoess and C. Lang, "Energy management for hypervisor-based virtual machines," *on Proceedings of the USENIX Annual*, 2007.

[7] D.C. Snowdon, E. Le Suer, S.M. Petters, and G. Heiser, "A platform for os- level power Management," *Eurosys '09*, 2009.

[8] H. Zeng, C.S. Ellis, A.R. Lebeck, and A. Vahdat, "ECOSystem: managing energy as a first class operating system resource," *ACM SIGPLAN Notices*, vol. 37, 2002, p. 123–132.

[9] A. Kansal, F. Zhao, and A.A. Bhattacharya, "Virtual Machine power Metering and Provisioning," *ACM Symposium on Cloud Computing (SOCC)*, 2010.

[10] Husain Bohra, Ata E, and Vipin Chaudhary. "VMeter: power modelling for virtualized clouds." *2010 IEEE International Symposium on Parallel & Distributed Processing, Workshops and Phd Forum (IPDPSW)* (April 2010): 1-8. http://ieeexplore.ieee.org/lpdocs/epic03/wrapper.htm?arnumber=5470907.

[11] Stallings, W.; , "SNMP and SNMPv2: the infrastructure for network management," *Communications Magazine, IEEE*, vol.36, no.3, pp.37-43, Mar 1998

[12] T. Heath, A. P. Centeno, P. George, L. Ramos, Y. Jaluria, and R. Biachini "Mercury and Freon: Temperature Emulation and Management in Server Systems". In International Conference on Architectural Support for Programming Languages and Operating Systems. Oct. 2006.

[13] W. L. Bircher and L.K. John. "Complete System power Estimation: A Trickle-Down Approach Based on Performance Events". In Internal Symposium on Performance Analysis of Systems and Software, April 2007.